\documentclass[twocolumn]{aastex62}

\usepackage[utf8]{inputenc}
\usepackage{graphicx}
\usepackage[percent]{overpic}
\usepackage{amsmath}   
\usepackage{enumitem}  

\graphicspath{{./}{figures/}}

\received{January 1, 2018}  
\revised{January 7, 2018}   
\accepted{\today}
\submitjournal{ApJ}

\shorttitle{Large-scale radio sources in NLSy1 galaxies}
\shortauthors{Umayal et al.}

\begin{document}

\title{Identification of Large-Scale ($>$100 kpc) Radio Jets in Narrow-Line Seyfert 1 Galaxies} 

\author[0009-0000-1823-5666]{S. Umayal}
\affiliation{Indian Institute of Astrophysics, Block II, Koramangala,
Bangalore 560 034, India}
\affiliation{National Remote Sensing Centre (NRSC), Indian Space Research Organisation (ISRO), Balanagar,
Hyderabad 500 037, India}

\correspondingauthor{Vaidehi S. Paliya}
\author[0000-0001-7774-5308]{Vaidehi S. Paliya}
\email{vaidehi.s.paliya@gmail.com}
\affiliation{Inter-University Centre for Astronomy and Astrophysics (IUCAA), SPPU Campus,
Pune 411 007, India}

\author[0000-0002-4464-8023]{D. J. Saikia}
\affiliation{Inter-University Centre for Astronomy and Astrophysics (IUCAA), SPPU Campus,
Pune 411 007, India}

\author[0000-0002-4998-1861]{C. S. Stalin}
\affiliation{Indian Institute of Astrophysics, Block II, Koramangala,
Bangalore 560 034, India}

\author[0000-0002-4024-956X]{S. Muneer}
\affiliation{Indian Institute of Astrophysics, Block II, Koramangala,
Bangalore 560 034, India}

\author[0009-0007-0745-9147]{Maheswar Gopinathan}
\affiliation{Indian Institute of Astrophysics, Block II, Koramangala,
Bangalore 560 034, India}

\begin{abstract}
Powerful, large-scale relativistic jets are usually associated with massive, old elliptical galaxies. This paradigm has recently been challenged by the identification of narrow-line Seyfert 1 (NLSy1) galaxies, thought to be young active galactic nuclei with low-mass black holes, harboring relativistic jets. Among them, sources hosting $>$100 kpc radio jets are extremely rare. 
Here, we report the discovery of large-scale, double-lobed radio structures in 33 NLSy1s with the projected linear size of at least 100 kpc from a recently published catalog of 22656 NLSy1 galaxies. These 33 include 29 confirmed double-lobed sources and 4 candidates whose radio structure require further study. We suggest that their low black hole masses are unlikely to be due to their small angles of inclination to the line of sight.
These enigmatic sources were identified by examining the radio observations taken with the Faint Images of the Radio Sky at Twenty centimeters, Very Large Array Sky Survey, Low Frequency Array, and Rapid ASKAP Continuum Survey. Among them, the NLSy1 source J1318+2626 ($z=0.62$) is found to host a radio jet with the projected linear size of 4.3 Mpc, making it the only NLSy1 galaxy hosting a Mpc-scale radio jet known as of today. 
We conclude that future observations of NLSy1 sources with the next-generation of sensitive telescopes may reveal a much larger population harboring large-scale jets, thus providing crucial clues on their origin, propagation, and interaction with the ambient environment.
\end{abstract}

\section{Introduction} \label{sec:intro}
Active galactic nuclei (AGN), whose luminosity can range up to $\sim$ 10$^{48}$ erg s$^{-1}$ \citep[e.g.][]{Padovani2017}, are believed to be powered by accretion of matter onto supermassive black holes \citep[SMBHs, 10$^6-10^{10}$ M$_{\odot}$; e.g., ][]{Woo&Urry2002}.
The observed emission over a wide range of wavelengths from AGN is mainly driven by the mass and spin of the SMBHs, and the mass accretion rate along with the orientation of the source axes to the line of sight \citep{Antonucci1993}.

Among the AGN are the narrow-line Seyfert 1 (NLSy1) galaxies, identified as a separate class about four decades ago. They are characterized by the relative narrowness of their broad emission lines with full width at half maximum of the broad H$\beta$ emission line less than 2000 km s$^{-1}$, weak [OIII] lines relative to H$\beta$ with [OIII]/H$\beta$ $<$ 3 and strong FeII multiplet emission \citep{Osterbrock1985,Goodrich1989}. However, it has been suggested that the narrow H$\beta$ lines in their spectra could be because of their orientation, with their broad line regions seen pole-on \citep[cf.][]{Decarli2008}. 

NLSy1 galaxies are believed to be powered by low mass black holes (10$^{6-8}$ M$_{\odot}$), having higher accretion rates, and predominantly hosted by spiral/disk galaxies undergoing rapid star-formation activity \citep[e.g.,][]{Mathur2000,Boroson1992,Jarvela2018,Varglund2022,Varglund2023,Sani2010,Caccianiga2015}. \cite{Mathur2000} suggested that NLSy1 galaxies could be young, rejuvenated AGN in the galaxy evolution scheme. However, some studies do point to NLSy1 galaxies having SMBH masses no different from their broad line counterparts, namely the broad-line Seyfert 1 (BLSy1) galaxies \citep{Calderone2013,Baldi2016,Liu2016,Rakshit&Stalin2017,Viswanath2019}. 

In X-rays, NLSy1 sources exhibit a steep 2$-$10 keV 
power-law spectrum with strong excess at energies below 2 keV \citep[e.g.,][]{Boller1996,Grunwald2023} and show large amplitude X-ray flux variability \citep[cf.][]{Boller1996,Leighly1999,Rani2017}. The hosts of NLSy1 galaxies are believed to be spiral/disks \citep[see, e.g.,][]{2003AJ....126.1690C,Jarvela2018,Varglund2022,Varglund2023}.  
However, elliptical hosts are also known, as for example in a few $\gamma$-ray emitting NLSy1 galaxies \citep[see, e.g.,][]{D'Ammando2017,D'Ammando2018}.

A small fraction of about 5\% of the known NLSy1 population are found to emit radio emission \citep{Zhou2006,Rakshit2017}. Among the radio-detected NLSy1s, about two-thirds are found to be radio loud \citep{Komossa2006,Zhou2006,2018MNRAS.480.1796S}. Interestingly, some NLSy1 galaxies exhibit strong radio emission, with compact cores of high brightness temperature,  strong radio variability, as well as flat or inverted radio spectra, pointing to these having relativistic jets similar to those of blazars \citep{Yuan2008,Gu2015,Lahteenmaki2017}. Furthermore, only a few NLSy1 galaxies are known to have extended radio structures ranging from few tens of kpc to around 100 kpc \citep{Doi2012,Rakshit2018,Doi2019,Vietri2022}. 

Interestingly, about two dozen radio-emitting NLSy1 galaxies are detected in the GeV band by the Large Area Telescope onboard the {\it Fermi} Gamma-ray Space Telescope \citep[e.g.,][]{Abdo2009,Paliya2018,Paliya2024b}. The broad band spectral energy distribution (SED) of these GeV emitting NLSy1 galaxies is similar to that of the flat spectrum radio quasar (FSRQ) category of blazars,  with the low-energy hump attributed to the synchrotron emission process and the inverse-Compton process contributing to the high energy hump \citep[cf.][]{Paliya2019b}. The broad-band SED as well as the parsec-scale radio polarimetric properties of the $\gamma$-emitting NLSy1 galaxies are similar to those of FSRQs \citep{Takamura2023}, in accordance with the scenario of GeV emitting NLSy1 galaxies being the low black hole mass counterparts to FSRQs \citep{Foschini2015,Paliya2018,Paliya2019b}. In a detailed study of two $\gamma-$ray emitting NLSy1 galaxies, \cite{Paliya2013} found these to be of lower luminosity than FSRQs, and suggested that their low-luminosity$-$low-frequency behavior could be due to their low black hole masses. These observations point to the unambiguous presence of relativistic jets in a minority of NLSy1 galaxies. They also challenge the notion that only AGN harboring massive black holes and residing in elliptical hosts can launch large-scale relativistic jets, contrary to AGN hosted by spiral/disk galaxies and powered by low mass black holes \citep[cf.][]{Laor2000}. Even though GeV $\gamma$-rays are detected in a handful of NLSy1 galaxies pointing to the presence of relativistic jets, such jets in NLSy1 galaxies are generally rare, and even if present, they are usually less than a few tens of kpc \citep{Berton2018}. However, large-scale, collimated  Faranoff Riley type II (FRII; \citealt{Fanaroff1974}) sources with sizes larger than 100 kpc have been reported in two NLSy1 galaxies namely SDSS J103024.95+551622.7 (116 kpc,\citealt{Rakshit2018}) and 6dFGS gJ035432.8$-$134008 (240 kpc,\citealt{Vietri2022}). The rarity of large double-lobed radio sources in NLSy1 galaxies, points to either the failure of these objects to launch large-scale radio jets or/and an underestimation of the numbers due to limitations of the radio surveys. 

With the availability of the largest sample of NLSy1 galaxies \citep{Paliya2024b} and a number of sensitive, high-resolution radio surveys including at low frequencies, we examine the occurrence and properties of large, double-lobed radio sources greater than about 100 kpc associated with NLSy1 galaxies in this paper. In Section~\ref{sec:data} we briefly describe the sample selection. The details of the radio surveys used and the methodology adopted to estimate their physical parameters are explained in Section~\ref{sec:catalog}. The results and discussion are presented in Section~\ref{sec:results}, and the concluding remarks are in Section~\ref{sec:conclusions}. Throughout, we used a flat cosmology with $H_0 = 70~{\rm km~s^{-1}~Mpc^{-1}}$ and $\Omega_{\rm M} = 0.3$.

\section{Sample Selection} \label{sec:data}
We have used the recently released catalog of NLSy1 galaxies by \cite{Paliya2024b}. This contains a total of 22656 NLSy1 galaxies with redshifts $\leq$0.9, selected from the Sloan Digital Sky Survey (SDSS) Data Release 17. This new catalog has more than twice the number of previously identified NLSy1 galaxies and supersedes the earlier catalogs published by \citet{Zhou2006} and \citet{Rakshit2017}. We cross-matched these NLSy1 galaxies with the Faint Images of the Radio Sky at Twenty Centimeters survey (FIRST; \citet{Becker1995}) catalog within a search radius of 5 arcsec. This led us to a sample of 730 NLSy1 galaxies that are detected in radio by FIRST survey. Details on the radio properties of all these sources will be published elsewhere (Umayal et al., in preparation).

\section{Multiwavelength Catalogs and Search Strategy} \label{sec:catalog}
In this work, we attempt to identify double-lobed radio structures with projected linear sizes of at least 100 kpc in the radio-detected NLSy1 galaxies, and study their properties. To achieve these objectives, we examined each of the 730 FIRST-detected NLSy1 galaxies, by overlaying their optical $r$-band images from the Panoramic Survey Telescope and Rapid Response System (Pan-STARRS; \citet{Chambers2016}), with the radio images covering a 10$\times$10 $\text{arcmin}^2$ region from FIRST, the Very Large Array Sky Survey (VLASS; \citet{Lacy2020}), the Rapid ASKAP Continuum Survey (RACS; \citet{McConnell2020}) at 887 MHz, the LOw Frequency ARray (LOFAR; \citet{vanHaarlem2013}) survey with both 6-arcsec and 20-arcsec resolutions, and NRAO VLA Sky Survey \citep[NVSS;][]{1998AJ....115.1693C}. In order to explore the possibility of diffuse extended emission which may be missed at higher frequencies with limited sensitivity, we also examined the sensitive, low-frequency LOFAR images over a region of 30$\times$30 $\text{arcmin}^2$. The details of the radio surveys are given in Table~\ref{table-1}.

To identify NLSy1s with projected linear sizes of at least 100 kpc, the radio contours of each of the 730 sources from the considered radio surveys were superposed on the r-band image from Pan-STARRS. The contours were plotted starting from the 3$\sigma$ level in powers of two, where $\sigma$ is the rms noise estimated locally. In addition, for an initial screening, circles of diameter 100 kpc were drawn around the target sources, which were derived from their respective redshifts. Sources with evidence of bipolar radio emission on opposite sides of the radio core or nucleus of the host galaxy, including significantly misaligned sources such as those reminiscent of Wide Angle Tailed (WAT) sources, were considered. Intrinsic misalignments may also  appear amplified when sources are inclined at small angles to the line of sight \citep[e.g.][]{KapahiSaikia1982}.

\begin{table}
    \caption{\textbf{Details of the radio surveys} used to identify NLSy1 galaxies with a projected linear size of at least 100 kpc.}
    \label{table-1}
    \centering
    \begin{tabular}{cccc}
        \hline
        Survey & Frequency & Angular & Sensitivity \\
        ~ & (GHz) &  resolution (\arcsec) &  (mJy/beam) \\ 
        \hline
        LOFAR 20\arcsec & 0.144 & 20 & 0.095\\
        LOFAR 6\arcsec & 0.144 & 6 & 0.083 \\ 
        RACS-low & 0.887 & 25 & 0.300 \\ 
        RACS-mid & 1.37 & 8.1-47.5 & 0.198 \\
        NVSS & 1.4 & 45 & 0.450 \\ 
        FIRST & 1.43 & 5.4 & 0.140 \\
        VLASS & 3.0 & 2.5 & 0.120 \\ 
        \hline
    \end{tabular}
    \begin{flushleft}
    {\setlength{\parskip}{0pt}
    \setlength{\parsep}{0pt}
    \footnotesize
    \hangindent=0.5em
    \hangafter=1
    \noindent
    }
    \end{flushleft}
\end{table}

For each candidate source whose size is likely to be at least 100 kpc, the largest angular sizes (LAS, in \arcsec) were accurately measured. For a radio lobe with a prominent hotspot at the outer edge, the angular separation was measured from the brightest pixel in the hotspot to the radio core or nucleus of the host galaxy. For a source with a diffuse outer lobe, the separation was measured from the farthest 3$\sigma$ brightness contour to the radio core. The separations of the two outer lobes from the radio core or nucleus of the host galaxy, estimated from an image where the overall structure is best represented, were added to derive the LAS for each source, and then converted to a projected linear size, D$_{\rm {proj}}$.

\begin{figure*}
    \centering
    \includegraphics[
        width=\textwidth,
        trim=20cm 17cm 20cm 20cm,
        clip
    ]{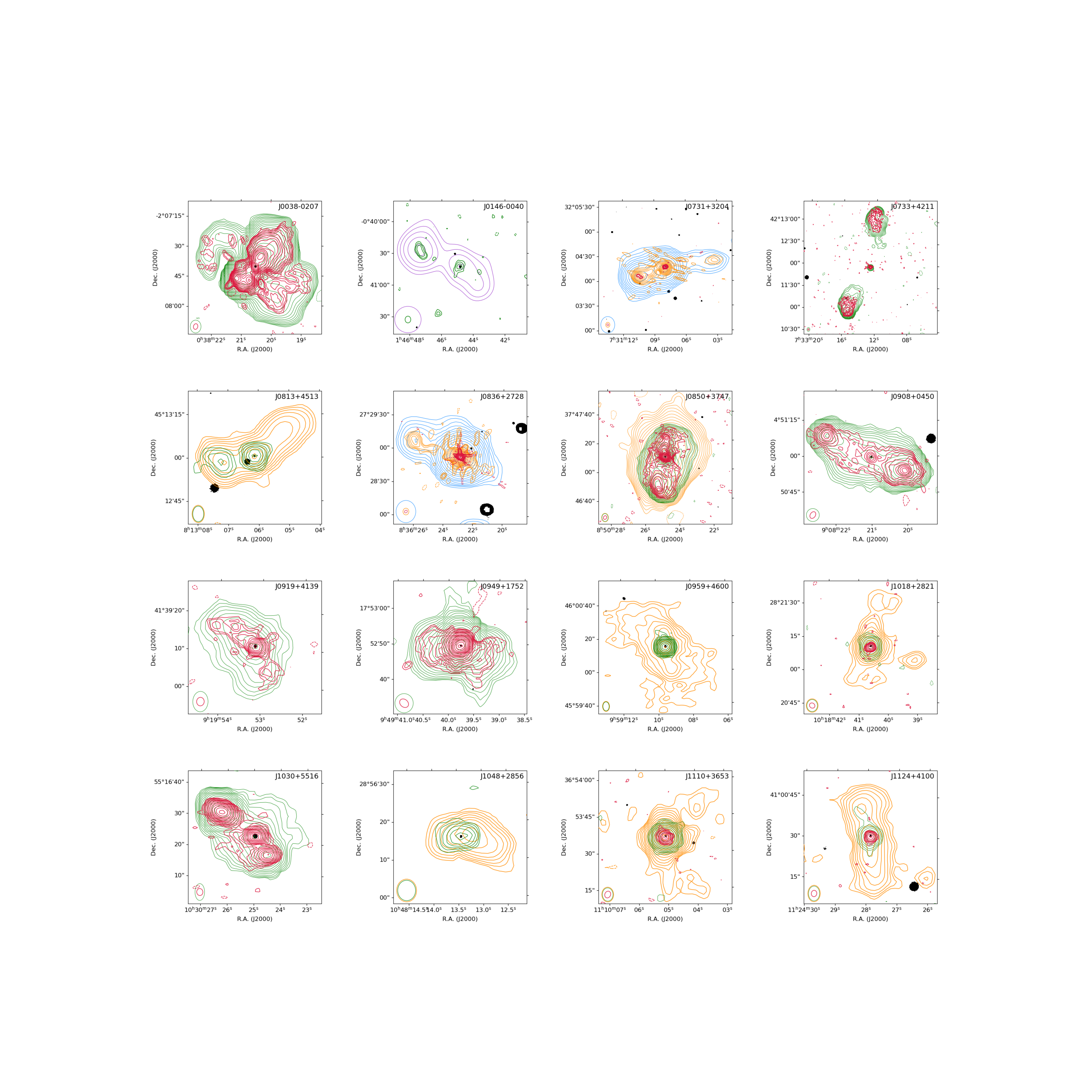}
    \caption{Total intensity maps of all the 34 NLSy1 sources, with projected sizes of at least 100 kpc. 
    The background grey images are the Pan-STARRS r-band images, while the contours are the radio images 
    from VLASS (red), FIRST (green), LOFAR 6$\arcsec$ resolution (orange), LOFAR 20$\arcsec$ resolution (blue) 
    and RACS-low (violet). The contour levels are at 3$\sigma \times [-(\sqrt{2}^m), (\sqrt{2})^n]$ with m=0 and n=[0, 1, 2, 3, 4, 5,...] mJy/beam. The dashed contours represent negative values. The lower left corner of each image shows the beams of 
    the different surveys. The names of the sources are given in the respective panels.}
    \label{morph-1}
\end{figure*}

\begin{figure*}
    \figurenum{1} 
    \centering
    \includegraphics[
        width=\textwidth,
        trim=20cm 26cm 20cm 30cm,
        clip
    ]{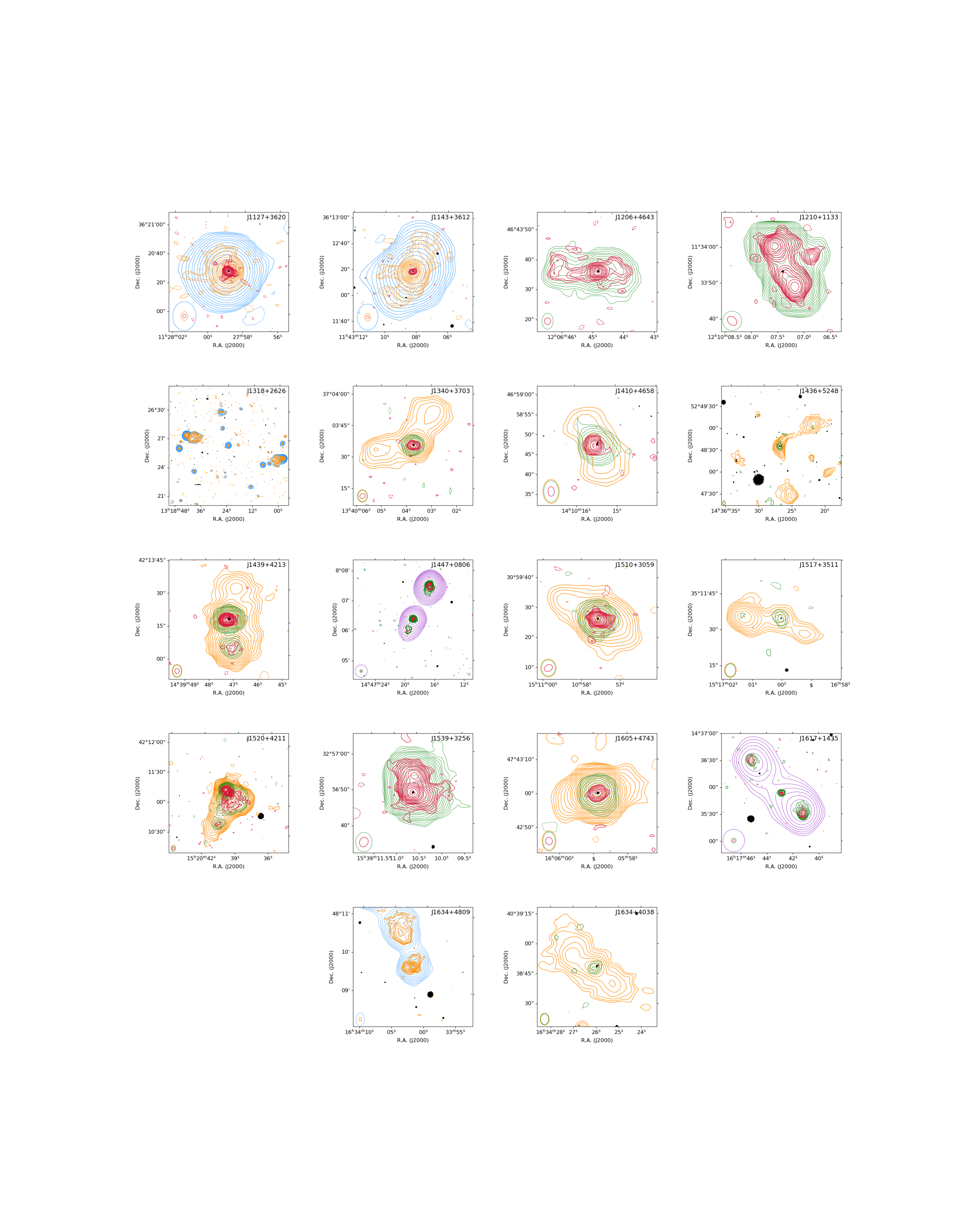}
    \caption{(Continued) Total intensity maps of NLSy1 sources.}
    \label{morph-1-contd}
\end{figure*}

\section{Results and Discussion} \label{sec:results}
A careful inspection of the cutout images taken from different radio surveys (Table~\ref{table-1}) has resulted in a sample of 34 NLSy1s with projected linear sizes of at least 100 kpc. Among them, J1030+5516 was reported earlier \citep{Rakshit2018}, while the remaining 33 are new identifications. For four of these, J1018+2821, J1048+2856, J1110+3653 and J1127+3620, radio images of higher sensitivity and/or resolution would be helpful to image the detailed structure. We have marked these as `candidates' in Table~\ref{table-2} with a superscript $^c$ after the source name. All 34 objects are listed in Table~\ref{table-2}, together with their physical parameters. The absolute $B$-band magnitudes (column 5, $\rm M_{B}$) and the masses of the SMBHs (column 6, $\rm M_{BH}$) are from \cite{Paliya2024b}. The radio morphological classification (column 9) is based on images that best represent the structure from which the LAS were estimated; the respective surveys used are listed in column 14.

\begin{table*}
    \caption{\textbf{Details of the NLSy1 galaxies with projected linear size of at least 100 kpc.} The columns are arranged as follows: 1. Source name, 2. R.A (J2000) in degrees, 3. Dec (J2000) in degrees, 4. Redshift z, 5. Absolute B magnitude M$_B$, 6. Mass of black hole (M$_{\rm SMBH}$/M$_\odot$) in logarithm scale, 7. Largest angular separation in arcsec, 8. Projected linear size in kpc, 9. Radio morphology which lists the Fanaroff-Riley class while X and WAT denote an X-shaped and a wide-angle tailed source respectively, 10. Core dominance parameter, 11. k$-$corrected extended luminosity $\rm{L_{ext}/WHz^{-1}}$ in logarithm scale at 150 MHz, 12. Corresponding jet power $\rm Q_{ext}/W$ in logarithm scale at the rest frequency of 150 MHz, 13. Radio spectral index $\alpha$, 14. Radio survey RS from which sizes have been measured (F: FIRST, L6 and L20: LOFAR with angular resolutions of 6 and 20 arcsec respectively, R: RACS and V: VLASS).}
    \label{table-2}
    \centering
    \begin{tabular}{lcccccrrcrcccl}
    \hline
    Source. ID & RA & DEC & z & M$_{B}$ & $\rm M_{BH}$ & LAS & $\rm D_{proj}$ & Morph. & $\rm C_D$ & $\rm{L_{ext}}$ & $\rm Q_{ext}$ & $\alpha$ & RS \\
    $[1]$ & $[2]$ & $[3]$ & $[4]$ & $[5]$ & $[6]$ & $[7]$ & $[8]$ & $[9]$ & $[10]$ & $[11]$ & $[12]$ & $[13]$ & $[14]$ \\
    \hline
    J0038$-$0207 & 9.5856 & -2.1279 & 0.2204 & -21.0 & 7.10 & 66.4 & 236.2 & X,FR I-II & -1.07 & 27.7 & 38.2 & -0.74 & F \\
    J0146$-$0040 & 26.6868 & -0.6787 & 0.0824 & -21.7 & 6.57 & 81.4 & 126.1 & FR I & -0.85 & 24.2 & 34.7 & -1.50$^{**}$ & R \\
    J0731+3204 & 112.7831 & 32.0712 & 0.7930 & -24.0 & 6.23 & 94.1 & 704.5 & FR II & -0.20 & 26.5 & 37.1 & -0.61 & L6 \\
    J0733+4211 & 113.3026 & 42.1991 & 0.4823 & -20.2 & 6.70 & 144.9 & 867.3 & FR II & -1.23 & 26.9 & 37.5 & -0.71 & V \\
    J0813+4513 & 123.2755 & 45.2167 & 0.8898 & -21.2 & 7.42 & 44.4 & 345.0 & WAT, FR I & -0.25 & 26.1 & 36.7 & -1.18 & L6 \\
    J0836+2728 & 129.0954 & 27.4813 & 0.7620 & -23.8 & 7.60 & 78.3 & 577.8 & FR I-II? & 0.54 & 26.3 & 36.8 & 0.02 & L6 \\
    J0850+3747 & 132.6030 & 37.7859 & 0.4073 & -20.9 & 7.14 & 37.8 & 205.3 & FR II & -0.38 & 27.2 & 37.8 & -0.60 & V \\
    J0908+0450 & 137.0876 & 4.8499 & 0.5244 & -21.6 & 6.40 & 35.9 & 224.9 & FR II & -1.57 & 27.3 & 37.8 & -0.85 & V \\
    J0919+4139 & 139.9715 & 41.6530 & 0.7749 & -24.0 & 7.35 & 19.5 & 144.9 & WAT?, FR I & -0.17 & 26.4 & 37.0 & -0.71 & V \\
    J0949+1752 & 147.4157 & 17.8804 & 0.6929 & -25.3 & 7.67 & 32.9 & 234.0 & WAT?, FR I & -0.02 & 27.4 & 37.9 & -0.44 & F \\
    J0959+4600 & 149.7896 & 46.0040 & 0.3989 & -21.4 & 6.96 & 77.8 & 417.3 & FR I & 0.56 & 25.2 & 35.7 & -0.14$^{*}$ & L6 \\
    J1018+2821$^c$ & 154.6691 & 28.3528 & 0.3841 & -21.4 & 6.21 & 44.7 & 234.5 & FR I & 0.29 & 24.9 & 35.4 & -0.61$^{*}$ & L6 \\
    J1030+5516 & 157.6040 & 55.2730 & 0.4350 & -25.0 & 7.82 & 20.6 & 116.1 & FR II & -0.79 & 26.7 & 37.2 & -0.65 & V \\
    J1048+2856$^c$ & 162.0560 & 28.9377 & 0.7906 & -23.8 & 7.27 & 26.0 & 194.2 & FR I & 0.27 & 25.4 & 35.9 & -0.52$^{*}$ & L6 \\
    J1110+3653$^c$ & 167.5210 & 36.8934 & 0.6300 & -22.2 & 7.24 & 59.5 & 407.0 & WAT?, FR I & 0.10 & 26.0 & 36.6 & -0.44 & L6 \\
    J1124+4100 & 171.1162 & 41.0084 & 0.8445 & -22.7 & 7.35 & 41.5 & 317.2 & FR I & 0.38 & 25.5 & 36.0 & -0.49$^{*}$ & L6 \\
    J1127+3620$^c$ & 171.9953 & 36.3412 & 0.8840 & -23.1 & 7.32 & 68.1 & 527.7 & FR I? & 0.62 & 26.2 & 36.7 & 0.31 & L20 \\
    J1143+3612 & 175.7843 & 36.2052 & 0.3994 & -20.9 & 6.25 & 77.1 & 413.9 & FR I & 0.43 & 25.3 & 35.8 & -0.89$^{*}$ & L20 \\
    J1206+4643 & 181.6870 & 46.7268 & 0.8420 & -23.1 & 7.30 & 26.0 & 198.5 & FR I & -0.12 & 26.7 & 37.2 & -0.64 & V \\
    J1210+1133 & 182.5309 & 11.5648 & 0.8620 & -23.3 & 7.46 & 13.0 & 100.0 & FR II & -1.09 & 27.7 & 38.3 & -0.79 & V \\
    J1318+2626 & 199.5976 & 26.4397 & 0.6234 & -23.1 & 7.46 & 638.7 & 4345.1 & FR II & -0.45 & 26.1 & 36.7 & -0.80$^{*}$ & L6 \\
    J1340+3703 & 205.0155 & 37.0599 & 0.7268 & -23.5 & 7.34 & 51.7 & 374.8 & WAT, FR I & 0.30 & 25.5 & 36.0 & -0.52$^{*}$ & L6 \\
    J1410+4658 & 212.5644 & 46.9798 & 0.6231 & -22.5 & 6.72 & 21.4 & 145.7 & FR I & 1.21 & 24.4 & 34.9 & -0.13 & L6 \\
    J1436+5248 & 219.1124 & 52.8111 & 0.4150 & -20.2 & 7.24 & 124.1 & 682.0 & WAT, FR I & -0.46 & 25.2 & 35.7 & -1.04$^{*}$ & L6 \\
    J1439+4213 & 219.9463 & 42.2215 & 0.4275 & -22.6 & 7.03 & 28.5 & 159.4 & FR II? & 0.61 & 25.4 & 35.9 & -0.31$^{*}$ & L6 \\
    J1447+0806 & 221.8287 & 8.1063 & 0.7923 & -24.9 & 6.91 & 105.7 & 791.3 & FR II & -0.32 & 27.3 & 37.9 & -0.65 & F \\
    J1510+3059 & 227.7399 & 30.9906 & 0.7900 & -23.6 & 7.19 & 35.1 & 262.0 & FR I? & 0.77 & 25.4 & 36.0 & -0.04$^{*}$ & L6 \\
    J1517+3511 & 229.2502 & 35.1930 & 0.6774 & -22.6 & 6.95 & 27.3 & 192.5 & FR I? & 0.11 & 25.1 & 35.6 & -0.98$^{**}$ & L6 \\
    J1520+4211 & 230.1654 & 42.1864 & 0.4850 & -22.8 & 7.63 & 86.6 & 519.9 & FR I & -0.06 & 26.6 & 37.1 & -0.38 & L6 \\
    J1539+3256 & 234.7942 & 32.9469 & 0.7250 & -24.6 & 7.51 & 16.7 & 121.2 & WAT?, FR I-II & -0.10 & 27.0 & 37.5 & -0.59 & V \\
    J1605+4743 & 241.4953 & 47.7167 & 0.9000 & -23.6 & 7.65 & 29.1 & 226.7 & FR I? & -0.20 & 26.5 & 37.0 & -0.91$^{*}$ & L6 \\
    J1617+1435 & 244.4286 & 14.5982 & 0.6573 & -23.2 & 7.38 & 84.9 & 591.2 & FR II & -1.04 & 26.6 & 37.2 & -1.13$^{*}$ & F \\
    J1634+4809 & 248.5081 & 48.1612 & 0.4949 & -22.6 & 7.21 & 109.2 & 662.9 & WAT, FR I? & -0.06 & 25.7 & 36.2 & -0.58$^{*}$ & L6 \\
    J1634+4038 & 248.6081 & 40.6468 & 0.4651 & -21.1 & 6.01 & 52.6 & 308.8 & FR I & -0.33 & 25.1 & 35.7 & -0.86$^{**}$ & L6 \\
    \hline
    \end{tabular}
    \begin{flushleft}
    {\setlength{\parskip}{0pt}
    \setlength{\parsep}{0pt}
    \footnotesize
    \hangindent=0.5em
    \hangafter=1
    \noindent
    \textbf{Note.} Spectral index values marked with an asterisk (*) are estimated using available archival data. Values with double asterisks (**) are two-point spectral indices due to limited data availability. $^c$ denotes candidate sources where radio images of higher resolution and/or sensitivity would help clarify its detailed structure.
    }
    \end{flushleft}
\end{table*}

Table~\ref{table-2} also lists the degree of core prominence, $\rm C_D$, defined as the ratio of core to extended flux density $\mathrm{F_{core}/F_{ext}}$ at a rest frame frequency of 3 GHz, in column 10. For each of the sources, we measured the radio core flux density ($\mathrm{F_{core}}$) from the high-resolution VLASS, else from FIRST images, bounded by their corresponding radio beams centered at the core. Similarly, the total flux density ($\rm{F_{total}}$) was from the low-resolution images, which was NVSS in most of the sources. $\rm{F_{total}}$ was estimated within the 3$\sigma$ level, which includes weak and diffuse radio emission from the radio lobes and excludes any possible unrelated sources in the vicinity. The $\rm{F_{total}}$ and $\mathrm{F_{ext} = F_{total} - F_{core}}$ were extrapolated to the rest frame frequency at 3 GHz with spectral indices $\rm{\alpha_{core}}$ = 0 and $\rm{\alpha_{ext} = -0.8}$, where the spectral index $\alpha$ is defined as $\mathrm{F_{\nu} \propto \nu ^ {\alpha}}$, where $\rm{F_{\nu}}$ is the flux density at frequency $\nu$. $\rm C_D$ is given by the expression \citep[][]{Paliya2024a}
\begin{equation} \label{eq-CD}
   {\rm C_D} = \log \left( \frac{F_{\text{core}}}{F_{\text{ext}}} \cdot (1 + z)^{\alpha_{\text{core}} - \alpha_{\text{ext}}} \right)
\end{equation}

The k-corrected extended luminosities in column 11 were calculated using the relation
\begin{equation} \label{eq-Lext}
    L_{\rm ext} = 4\pi D_{L}^{2} F_{\text{ext}} (1+z)^{-(1+\alpha_{\text{ext}})}.
\end{equation}
Then the extended kinetic jet power (column 12) of the sources was calculated using the following relation (\citet{Hardcastle2018})
\begin{equation} \label{eq-Qext}
    L_{150}^{\rm ext} = 3 \times 10^{27} {\frac{Q_{\text{jet}}}{10^{38} W}}
\end{equation}

The integrated radio spectral indices ($\alpha$) of the sources (column 13) are from \cite{Stein2021}, when available. Otherwise, they have been evaluated from a linear least-squares fit to the available data (indicated by a single asterisk) or in three cases without adequate data the two-point spectral index values have been listed and indicated by double asterisks. 

The FIRST, VLASS, LOFAR and RACS contours of all these 34 sources, wherever available, superposed on the r$-$band images from Pan$-$STARRS are shown in Fig.~\ref{morph-1}. 

The detection of extended radio emission on the $>100$-kpc scale in NLSy1 galaxies has been relatively rare. Until now, only two such sources, J1030+5516 and J0354$-$1340 have been reported with projected sizes exceeding 100 kpc \citep[][]{Rakshit2018,Vietri2022}. Through a careful study of 730 NLSy1 sources detected in the FIRST survey from the largest compilation of 22,656 NLSy1 galaxies, we have identified 34 NLSy1 galaxies (including J1030+5516 reported by \citealt{Rakshit2018}) with $\mathrm{D_{\text{proj}} > 100}$ kpc. J0354$-$1340 \citep{Vietri2022} does not appear in our list as it is outside SDSS coverage area, hence absent from the NLSy1 catalog used in this work. High-resolution observations may resolve out the diffuse, extended emission, but could be useful for detecting extended emission on small scales. For example, of the 34 sources in our sample, three (J0146$-$0040, J1110+3653 and J1634+4809) have been observed by \cite{Berton2018} with sub-arcsec resolution with the VLA A-configuration at 5 GHz. In all three cases, the extended emission has been resolved out.

The large increase is likely due to the following reasons.
\begin{enumerate}
    \item The use of a number of sensitive radio surveys, especially at low frequencies have facilitated the detection of diffuse extended emission. For example, the LOFAR observations were critical in detecting the diffuse extended emission and determining their projected linear sizes reliably in over 50\% of the sources in our sample. In many of these, the extended emission would have been missed without LOFAR observations (e.g., J0959+4600). 
    
    \item In the recent catalog of NLSy1s by \cite{Paliya2024b} the redshift range has been extended to 0.9 by using the results from the Baryon Oscillation Spectroscopic Survey (BOSS) spectrographs. The BOSS spectrographs have a larger wavelength coverage than SDSS. Six of our sources have a redshift $>$0.8, the limit used for samples based on the SDSS spectrographs \cite[c.f.][]{Rakshit2017,Paliya2024b}.
    
\end{enumerate}

\subsection{Radio structure and sizes}
The Fanaroff-Riley classes of these sources are listed in Column 9 of Table~\ref{table-2}. Among these, 10 ($\sim$30\%) are FR~II doubles, 21 ($\sim$60\%) are FR~Is, and 3 ($\sim$10\%) have been classified as FR~I–IIs. A $'?'$ has been put against those where better images would help confirm their classification. J0038$-$0207 exhibits an X-shaped structure while 8 have structures reminiscent of wide-angle tailed sources.

\cite{Doi2012} noted that the majority of NLSy1s known to have kpc-scale structure at that time were FR~IIs, unlike what we find for our sample. The detection of a large number of FR~I sources has been possible largely due to the use of sensitive low-frequency radio surveys, such as LOFAR.

Their projected linear sizes D$_{\rm {proj}}$ range from 100 kpc to 4345 kpc, with a median value of 285 kpc. The largest one J1318+2626 has been listed as a giant radio galaxy (GRG) by \cite{Andernach2025}, and is the only NLSy1 hosting Mpc-scale jet known to date. A GRG is defined to be one with a projected linear size $\mathrm {D_{proj}>700}$ kpc \citep[see][for a review]{Dabhade2023}. Our sample contains three other GRGs, J0731+3204, J0733+4211, and J1447+0806  with $\mathrm {D_{proj}}$ of 704.5, 867.3 and 791.3 kpc respectively. The log of mass of the SMBHs in units of M$_\odot$ of these three GRGs are 6.23, 6.70 and 6.91 respectively, while that of J1318+2626 is 7.46. The presence of such gigantic structures originating from relatively low mass SMBHs hosted in NLSy1 galaxies poses interesting challenges in understanding the formation and launching of such large-scale jets.

\begin{figure*}
    \centering
    \begin{minipage}[t]{0.32\textwidth}
        \centering
        \begin{overpic}[width=\linewidth]
        {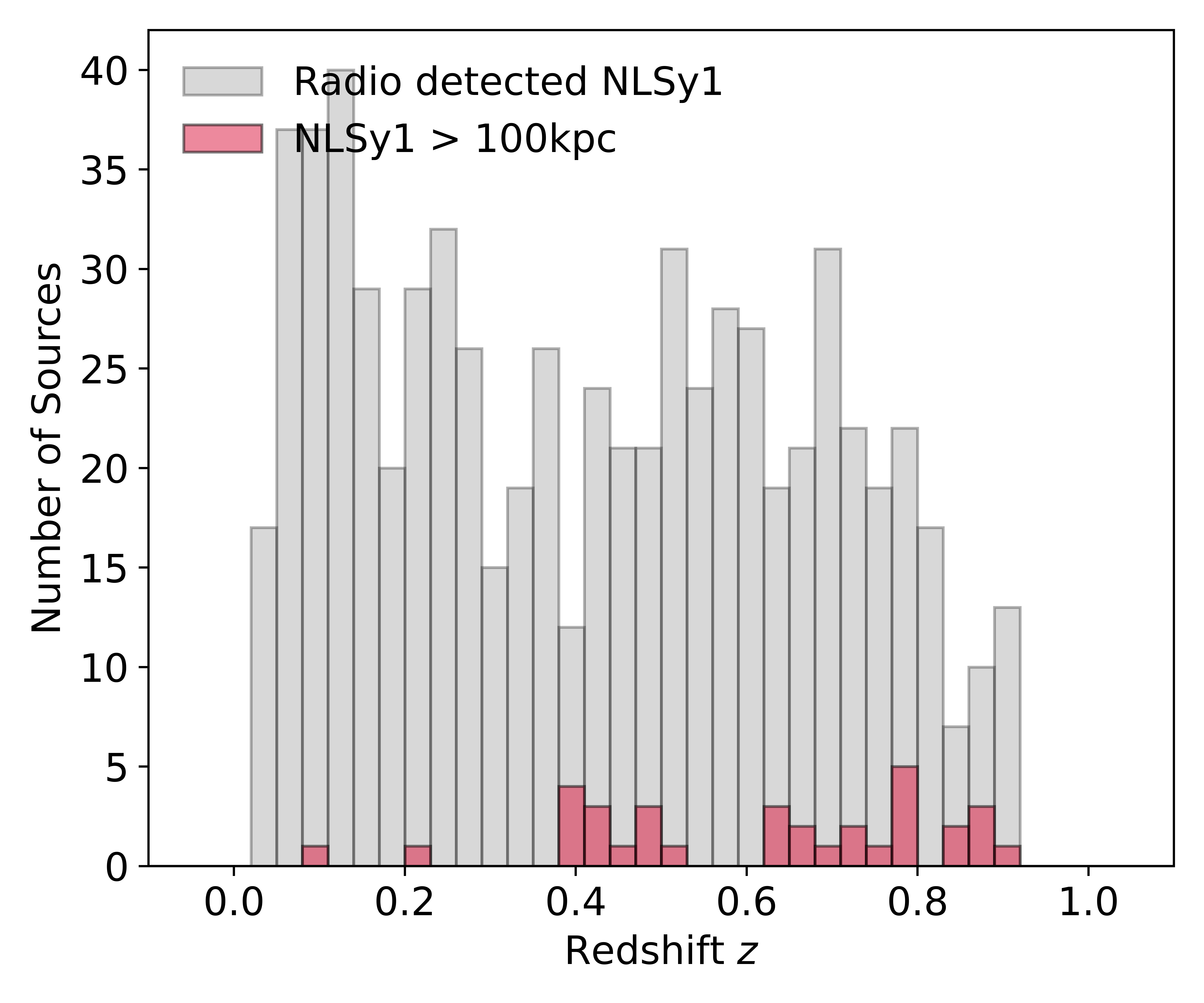}
            \put(13,14){\tiny (a)}
        \end{overpic}
    \end{minipage}
    \hfill
    \begin{minipage}[t]{0.32\textwidth}
        \centering
        \begin{overpic}[width=\linewidth]
        {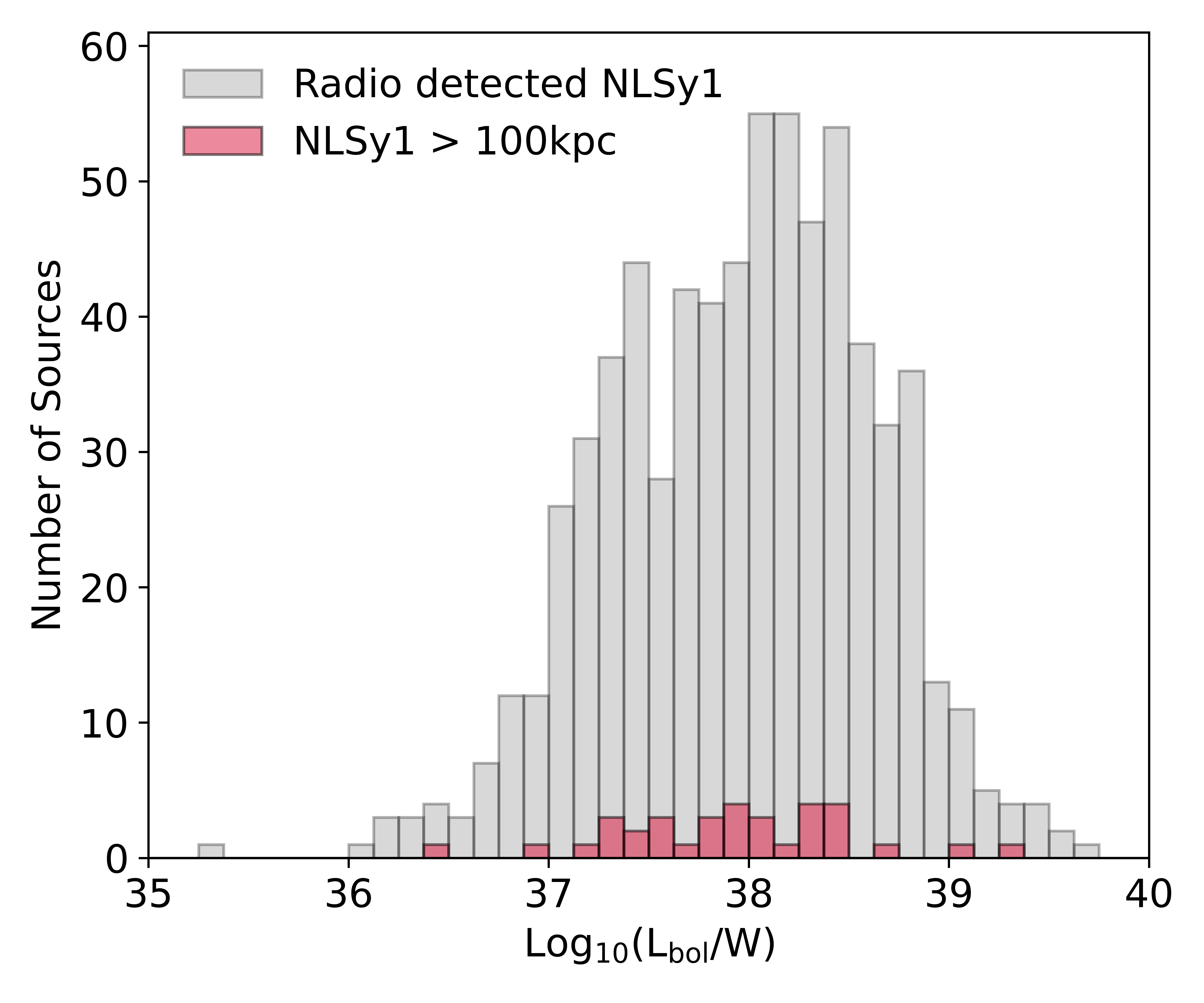}  
            \put(13,14){\tiny (b)}
        \end{overpic}
    \end{minipage}
    \hfill
    \begin{minipage}[t]{0.32\textwidth}
        \centering
        \begin{overpic}[width=\linewidth]
        {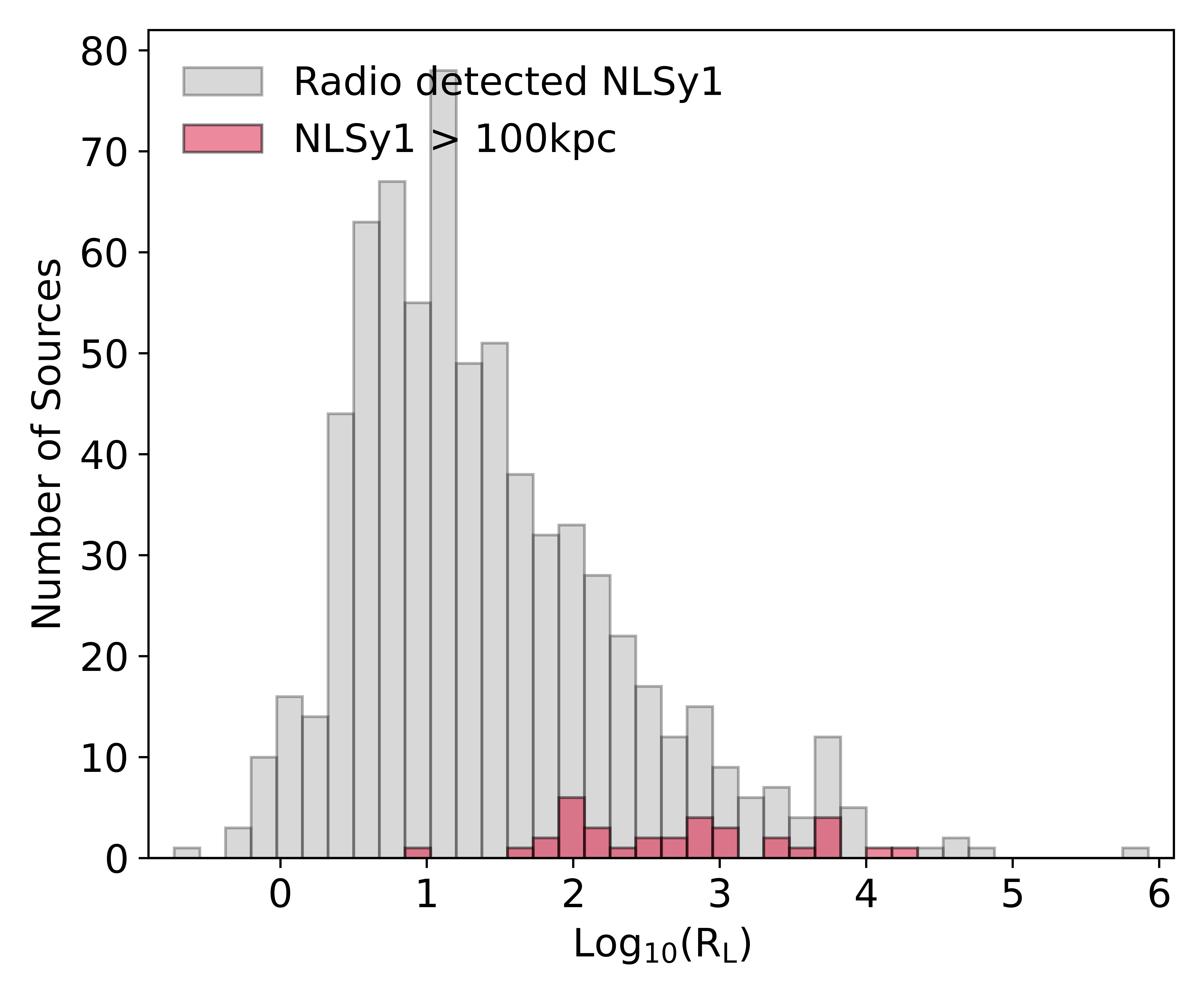}
            \put(14,14){\tiny (c)}
        \end{overpic}
    \end{minipage}
    \hfill
    \vspace{-0.1cm}
    \begin{minipage}[t]{0.32\textwidth}
        \centering
        \begin{overpic}[width=\linewidth]
        {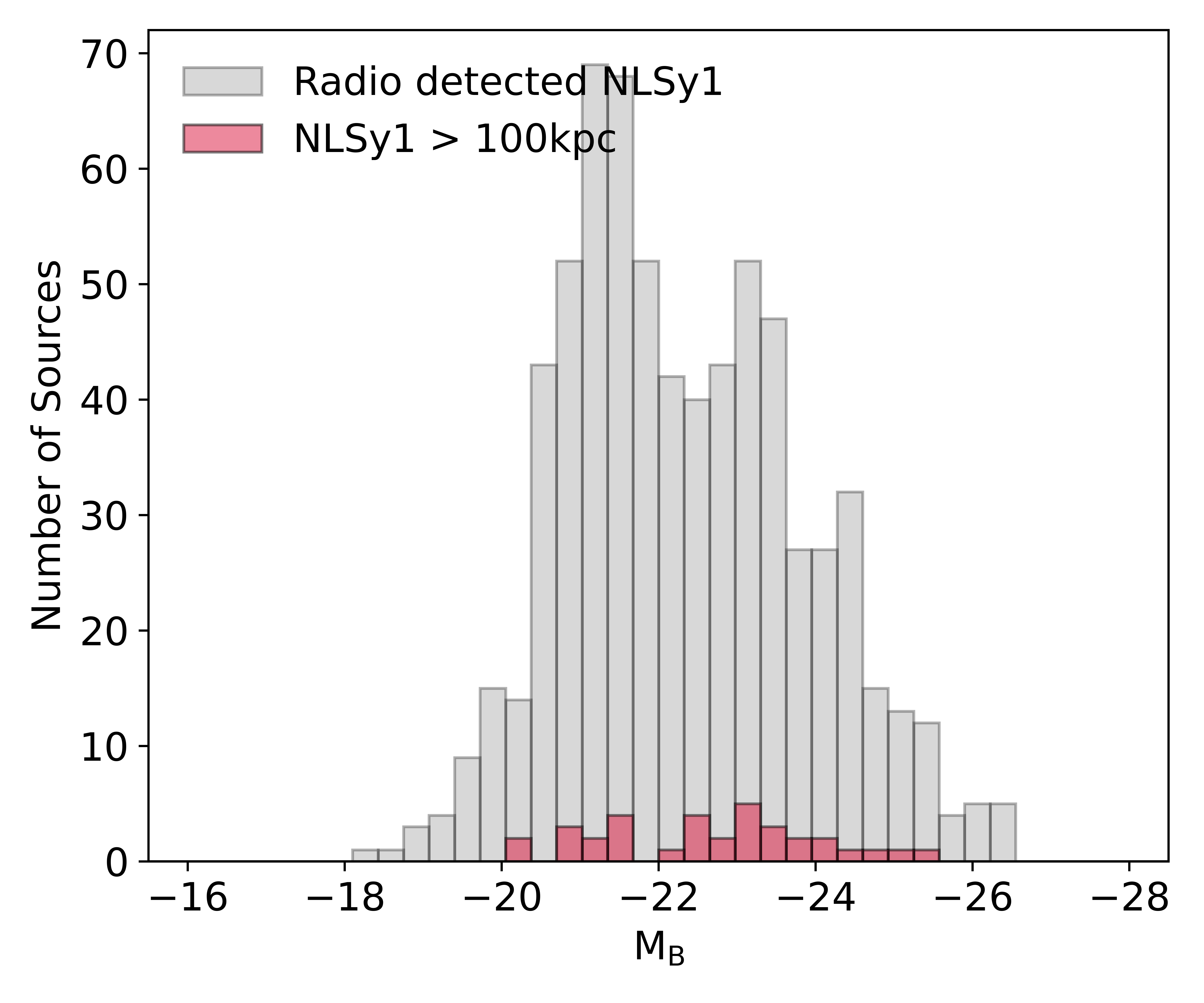}
            \put(13,14){\tiny (d)}
        \end{overpic}
    \end{minipage}
    \hfill
    \begin{minipage}[t]{0.32\textwidth}
        \centering
        \begin{overpic}[width=\linewidth]
        {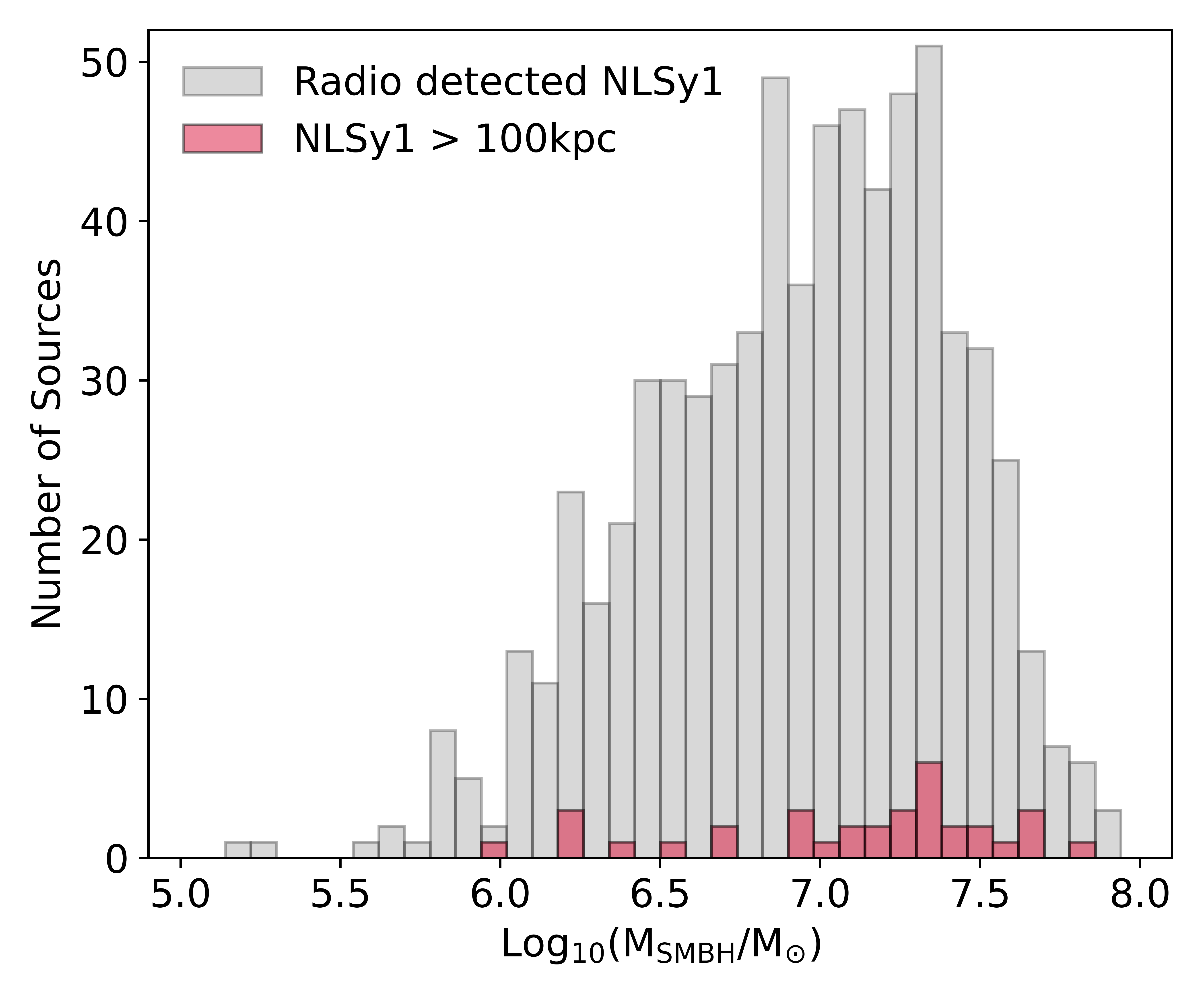}
            \put(13,14){\tiny (e)}
        \end{overpic}
    \end{minipage}
    \hfill
    \begin{minipage}[t]{0.32\textwidth}
        \centering
        \begin{overpic}[width=\linewidth]
        {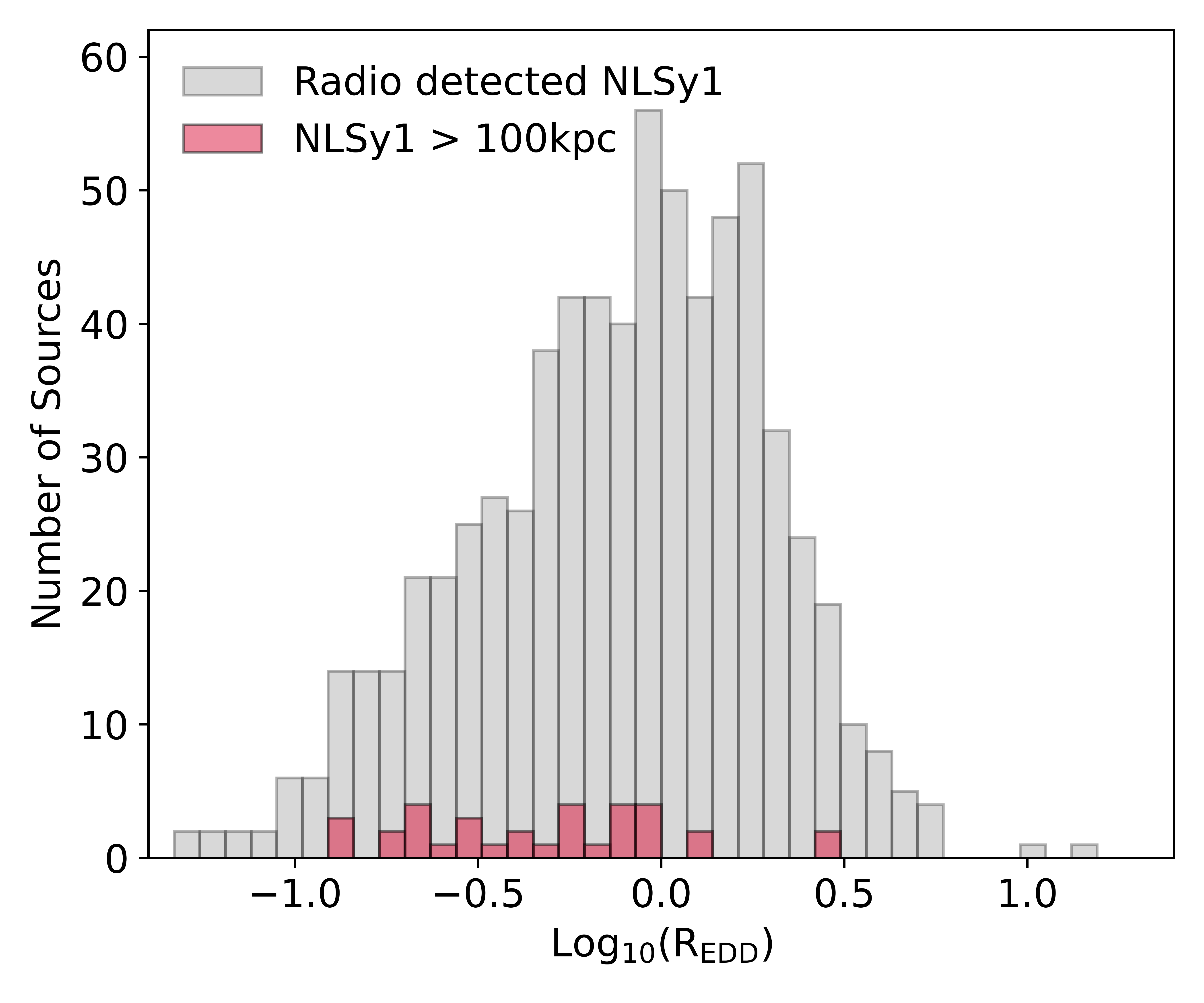}
            \put(15,14){\tiny (f)}
        \end{overpic}
    \end{minipage}
    \hfill
    \vspace{-0.25cm}    
    \caption{Distributions of (a) redshift z, (b) Bolometric luminosity and (c) radio loudness parameter are shown in the upper panels; while those of (d) absolute B magnitude, (e) mass of BHs and (f) Eddington ratio are shown in the bottom panels.
    The radio-detected NLSy1 sources plotted in gray do not include the $>$100~kpc NLSy1s.}
    \label{distributions}
\end{figure*}

\subsection{Comparison with radio-detected NLSy1s}
Here we briefly compare some of the properties of the NLSy1 sources $>$100 kpc, with other radio-detected NLSy1 sources, referred to as the radio-detected NLSy1 sample. We wish to explore any characteristics of the NLSy1 $>$100 kpc sources that enable them to launch these large-scale jets.

The redshift distributions of the two samples are shown in the top left panel of Fig.~\ref{distributions}. The median values are 0.42 and 0.64 for the radio-detected NLSy1 and the NLSy1 $>$100 kpc sources respectively, suggesting that the latter are seen at higher redshifts. A Kolmogorov-Smirnov test shows the two distributions to be significantly different with a p-value $<0.01$. This is likely due to the latest radio surveys being able to capture the faint, diffuse emission originating from these more distant objects. The bolometric luminosity distributions of the two samples (Fig.~\ref{distributions}, top middle panel) show that these are similar. The logarithmic median values, in Watt, are 38.02 and 37.95 for the radio-detected and the NLSy1 $>$100 kpc samples respectively. A Kolmogorov-Smirnov test yields a p-value of 0.43 showing that the two distributions are similar. A comparison of the radio-loudness parameter $\rm R_L$ shows that the NLSy1 $>$100 kpc have significantly higher values with a median at 525, compared with 15 for radio-detected sources (Fig.~\ref{distributions}, top right panel). A Kolmogorov-Smirnov test shows the two distributions to be significantly different with a p-value $<0.01$. In fact, a majority of the $>$100 kpc NLSy1 sources (28/34, $\sim$80\%) have a high $\rm R_L > 100$, and all the sources are radio loud with the exception of J0146$-$0040, which has a value ($\rm R_L = 7.1$) slightly lower than the threshold of $\rm R_L = 10$ \citep{Kellermann1989}, primarily because of it being extremely faint in the FIRST survey data.

The median values of the absolute B-band magnitude $\rm M_B$ values (Fig.~\ref{distributions}, bottom left panel) are $-22.17$ and $-22.75$ for the radio-detected and $>$100 kpc NLSy1 sources respectively. The $>$100 kpc sources tend to be only marginally more luminous.  A Kolmogorov-Smirnov test yields a p-value of 0.08, consistent with only a marginal difference.

The $\rm M_{BH}$ distributions (Fig.~\ref{distributions}, bottom middle panel) show that the median values of logarithmic $\rm M_{BH}$  are 6.99 and 7.24 for the radio-detected and $>$100 kpc NLSy1s respectively. The ones hosting $>$100 kpc radio structure appear to be more massive. In this case, a Kolmogorov-Smirnov test yields a p-value of 0.027, suggesting that the two distributions are different. The logarithmic $\rm M_{BH}$ distribution for the radio undetected NLSy1s is similar to that of the radio-detected ones, with a median value of 6.98, both being smaller than for the $>$100 kpc NLSy1s.

The Eddington ratio $\rm R_{EDD}$ distributions (Fig.~\ref{distributions}, bottom right panel) show that the median values are $-$0.07 and $-$0.31 for the radio-detected and the $>$100 kpc NLSy1s respectively. A Kolmogorov-Smirnov test shows the two distributions to be significantly different with a p-value of $<$0.01. The lower value for the $>$100 kpc NLSy1s could be largely due to their higher black hole masses.

To summarize, the $>$100 kpc NLSy1s in our sample tend to be at higher redshifts compared with the radio-detected ones. The $>$100 kpc NLSy1s are nearly all radio loud, with over 80 per cent having a value $\rm R_L > 100$. Therefore, choosing the NLSy1s with $\rm R_L > 100$ increases the probability of selecting those with large-scale jets. Although the $>$100 kpc NLSy1s tend to have a slightly higher black hole mass, this alone does not seem to be the sole requirement as there are GRGs among the least massive ones. In addition, spin of the black hole and availability of fuel for accretion over a long time scale may also possibly be playing a role.

\subsection{Orientation and core dominance}
The core dominance parameter ($\rm C_D$) has been used as a statistical measure of the orientation of a source or jet axis since the early days of the unification scheme for AGN \citep[e.g.][]{OrrBrowne1982,KapahiSaikia1982}. As the viewing angle decreases, beamed emission from the core dominates due to Doppler boosting, while the extended lobe emission is largely isotropic. Attempts have also been made to infer the orientation angle from their symmetry parameters, such as the separation ratio of the oppositely directed hotspots, and then estimate their deprojected sizes  \citep[e.g.][]{Rakshit2018}. If one assumes that the jets and the external environment are both symmetric, the receding hotspot is seen at an earlier time due to light travel time across the source, and appears closer to the core, while the approaching one appears farther. Also the approaching one is expected to be brighter due to mild relativistic beaming of the hotspots. However, there is increasing evidence from symmetry parameters \citep[][for a review]{Saikia2003,Saikia2022} as well as distributions of line-emitting gas \citep{McCarthy1991} in powerful radio galaxies that sources are not intrinsically symmetric, making it difficult to infer orientation angles from symmetry parameters. In our sample of $>$100 kpc NLSy1 sources too, there are several examples, such as J0731+3204, J0733+4211, J0908+0450, J1439+4213 and J1617+1435, all of which are FRII sources, where the nearer component is significantly brighter, underlining the importance of intrinsic asymmetries in the environment.

Traditionally the arm-length ratio distribution for a sample of high-luminosity FRII sources, where the hotspots are well defined, has been used to estimate the velocity of advancement of the hotspots, assuming sources to be intrinsically symmetric and randomly oriented in the sky. Such an exercise is not very meaningful for FRI sources with diffuse outer lobes, often residing in a rich, complex cluster environment. \cite{Scheuer1995} critically examined the hotspot velocity estimate from arm-length ratios for luminous radio sources and suggested that the maximum velocity of advancement for the high-luminosity radio galaxies is likely to be $<$0.15c. The median velocity of advancement of hotspots in compact steep-spectrum radio sources from high-resolution milli-arscsec scale observations of its motion is also $\sim$0.1c \citep[cf.][]{ODea2021}. The NLSy1s are unlikely to have a larger velocity of advancement than these luminous radio galaxies and quasars with powerful jets. In order to explore whether reliable estimates of orientation angles for our sample of NLSy1 galaxies are possible, we consider the ten sources with an FRII structure. Of these, for five of the sources listed above, the nearer component is significantly brighter showing that an asymmetric environment is playing a dominant role. For the remaining five, adopting a velocity of 0.15c as suggested by \cite{Scheuer1995}, only one source, J1318+2626, yields a consistent angle of orientation of about 32$^\circ$. Before inferring orientation angles from symmetry parameters, one needs a study of the environments of these sources from optical and X-ray observations.

\begin{figure}[t]
  \noindent 
  \hspace{0.16cm}\includegraphics[scale=0.53]{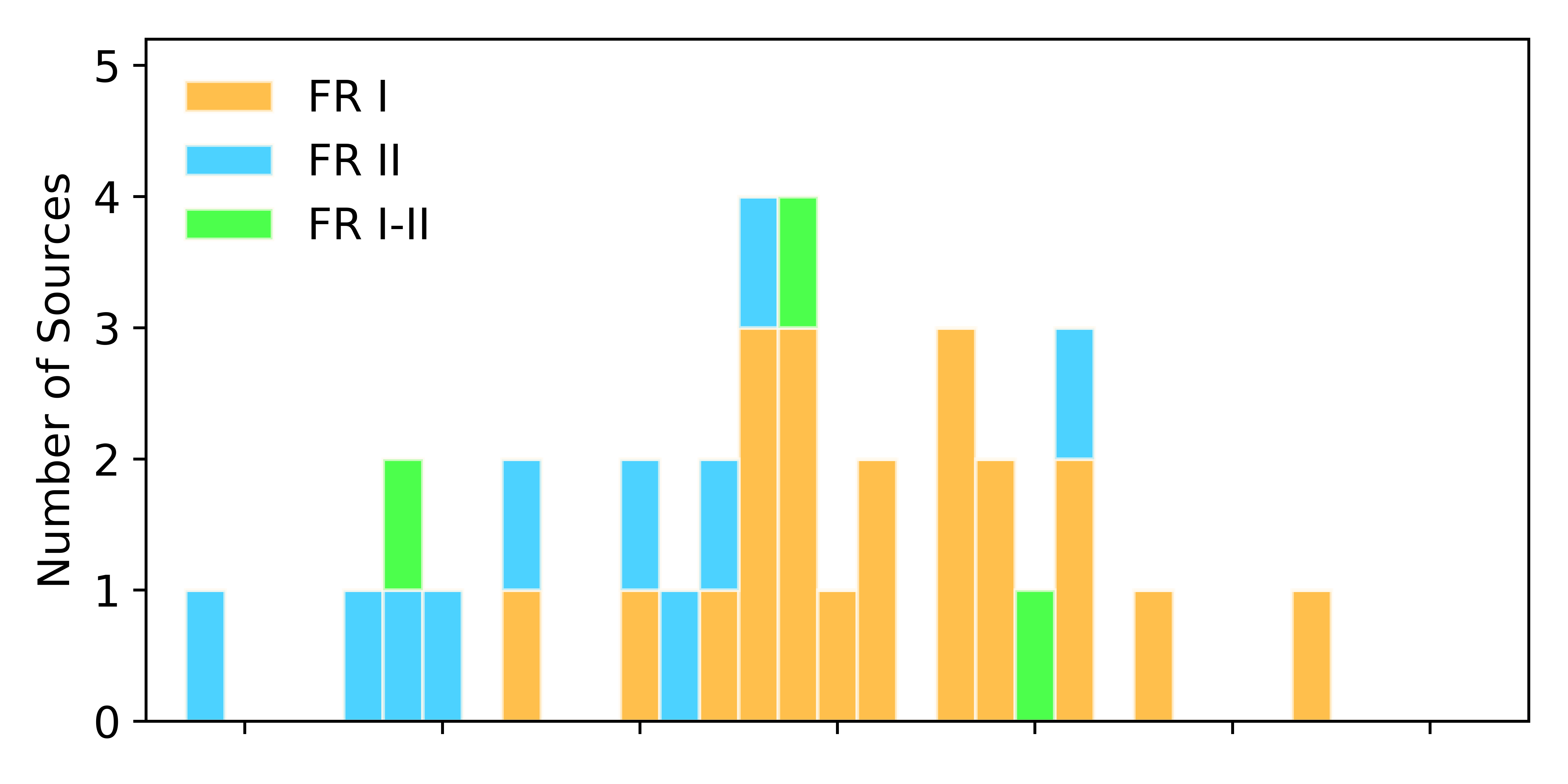}\vspace{-0.25cm}
  \includegraphics[scale=0.54]{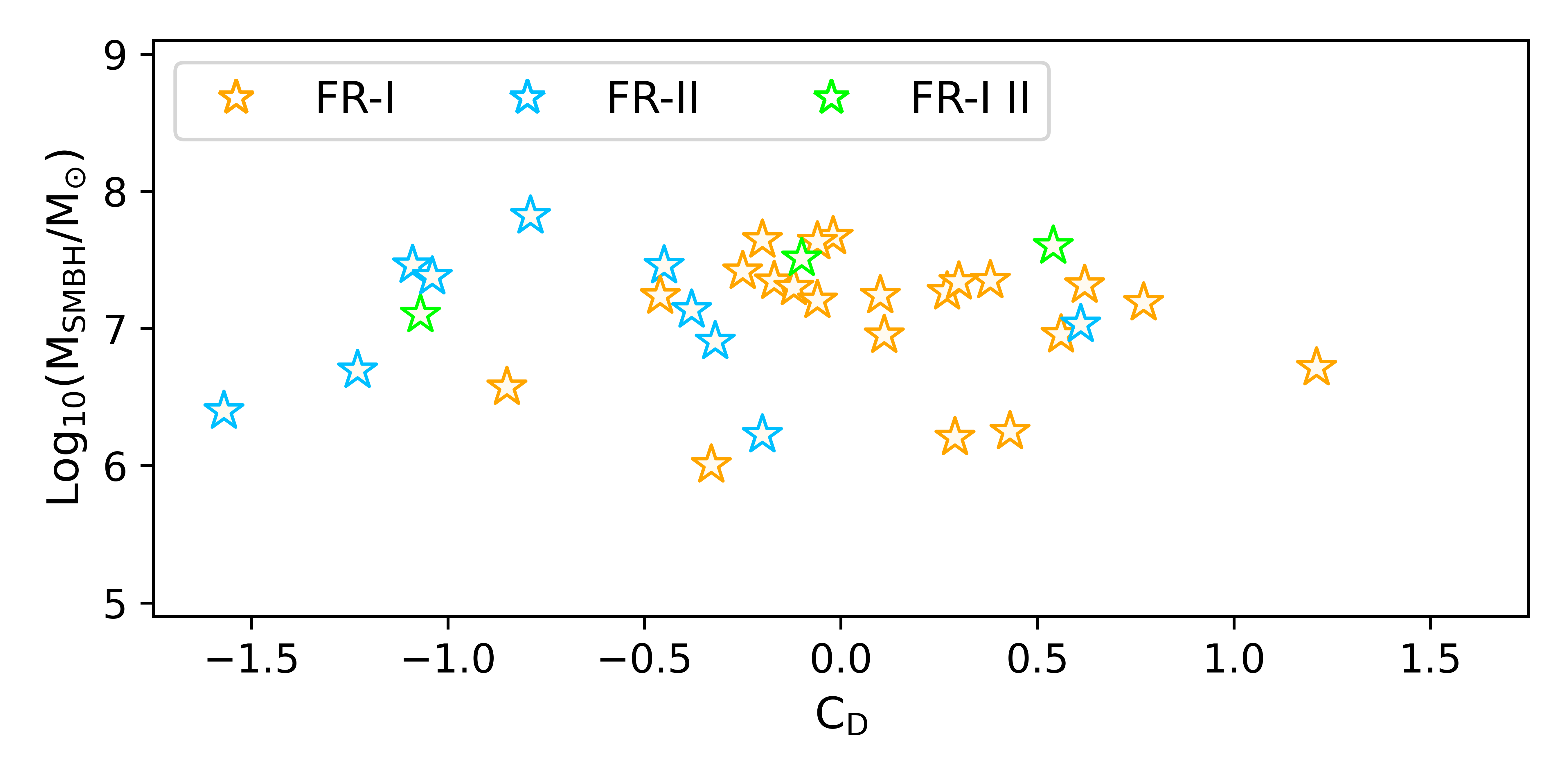}
  \vspace{0cm}
  \caption{Upper: Histogram of the estimated core dominance of the NLSy1s listed in Table~\ref{table-2}. 
  Lower: Relation between core dominance and black hole mass.}
  \label{CD_histogram}
\end{figure}

We estimated $\rm C_D$ using Equation~\ref{eq-CD}, obtaining values ranging from $-1.6$ to $1.2$, with a median value of $-$0.1 (Fig.~\ref{CD_histogram}). The FR~I sources tend to have higher values of $\rm C_D$, with a median value of about 0.1, compared with the FR~II sources which have a median value of about $\sim-0.6$. These results are consistent with earlier studies. In a study of B2 radio galaxies, most of which were lower-luminosity FR~I sources, along with the more luminous 3CR sources, \cite{deRuiter1990} showed that the core dominance decreased systematically with luminosity over about five decades. More recently, a similar trend was also seen in the sample of $\gamma$-ray emitting misaligned AGN \citep{Paliya2024a}. 

A small fraction of NLSy1 galaxies are known to exhibit blazar-like characteristics \citep[e.g.][]{Yuan2008,Gu2015,Lahteenmaki2017}, and about two dozen NLSy1 galaxies are known to be $\gamma-$ray emitters \citep{Paliya2019,Paliya2024b}. From our sample, there are four sources (J0836+2728, J0949+1752, J1127+3620 and J1520+4211) which appear in the systematically compiled catalog of blazars by \cite{Massaro2015}. These all have prominent cores with values of $\rm C_D$ ranging from $\sim -$0.1 to 0.6, and flat radio spectral indices with three of these having an inverted spectrum. Of these, J0949+1752, J1127+3620 and J1520+4211 appear in the most recent list of $\gamma-$ray emitting NLSy1s \citep{Paliya2024b}. The only other $\gamma-$ray emitting NLSy1 listed by \cite{Paliya2024b} is the X-shaped source J0038$-$0207, which has a relatively weak core with $\rm C_D=-$1.1 and a steep radio spectral index of $-$0.7. 
More recently, \cite{Gabanyi2025} have suggested J0959+4600 to be associated with a $\gamma-$ray source. It has a dominant core with $\rm C_D$=0.6 and a flat radio spectral index of $-$0.1. Their core dominance and flat spectral indices are consistent with these sources being inclined at small angles to the line of sight.

There have been suggestions that the small black hole masses estimated for NLSy1s could be due to a geometrical effect. If the broad-line region has a disk-like geometry, and its axis is inclined at a small angle to the line of sight, the velocity estimate and hence the black hole mass would be lower \citep[e.g.][]{Decarli2008}. One can examine this further using $\rm C_D$ as an indicator of the orientation of the source. In this case one would expect sources inclined at small angles to the line of sight to have higher values of $\rm C_D$ and smaller black masses. In the lower panel of Fig.~\ref{CD_histogram}, we plot $\rm C_D$ against the black hole mass for our sample of sources and find no significant trend. The median value of log of black hole mass in units of solar mass for the blazars listed earlier, which are expected to be inclined at small angles to the line of sight, is $\sim$7.6. Including the $\gamma-$ray emitting NLSy1s, the value is $\sim$7.3. All of these but one (J0959+4600 with a value of 6.96) have a value $>$7.  The GRGs, on the other hand, are generally expected to be at large angles to the line of sight. All the four GRGs in our sample have an FR~II structure and relatively weak cores with values of $\rm C_D$ ranging from $-$1.2 to $-$0.2. The mass of the black holes for three of the GRGs is less than 10$^{7} M_\odot$, with the largest one J1318+2626 having a logarithmic value of 7.46 (in $M_\odot$). These results suggest that orientation may not be playing a dominant role in determination of black hole mass of the NLSy1s. In this context it is also relevant to note that \citet[][]{Berton2021} summarized several arguments to rule out the possibility that NLSy1s are powered by black holes of high mass.

\begin{figure}[t]
  \noindent 
  \includegraphics[scale=0.41]{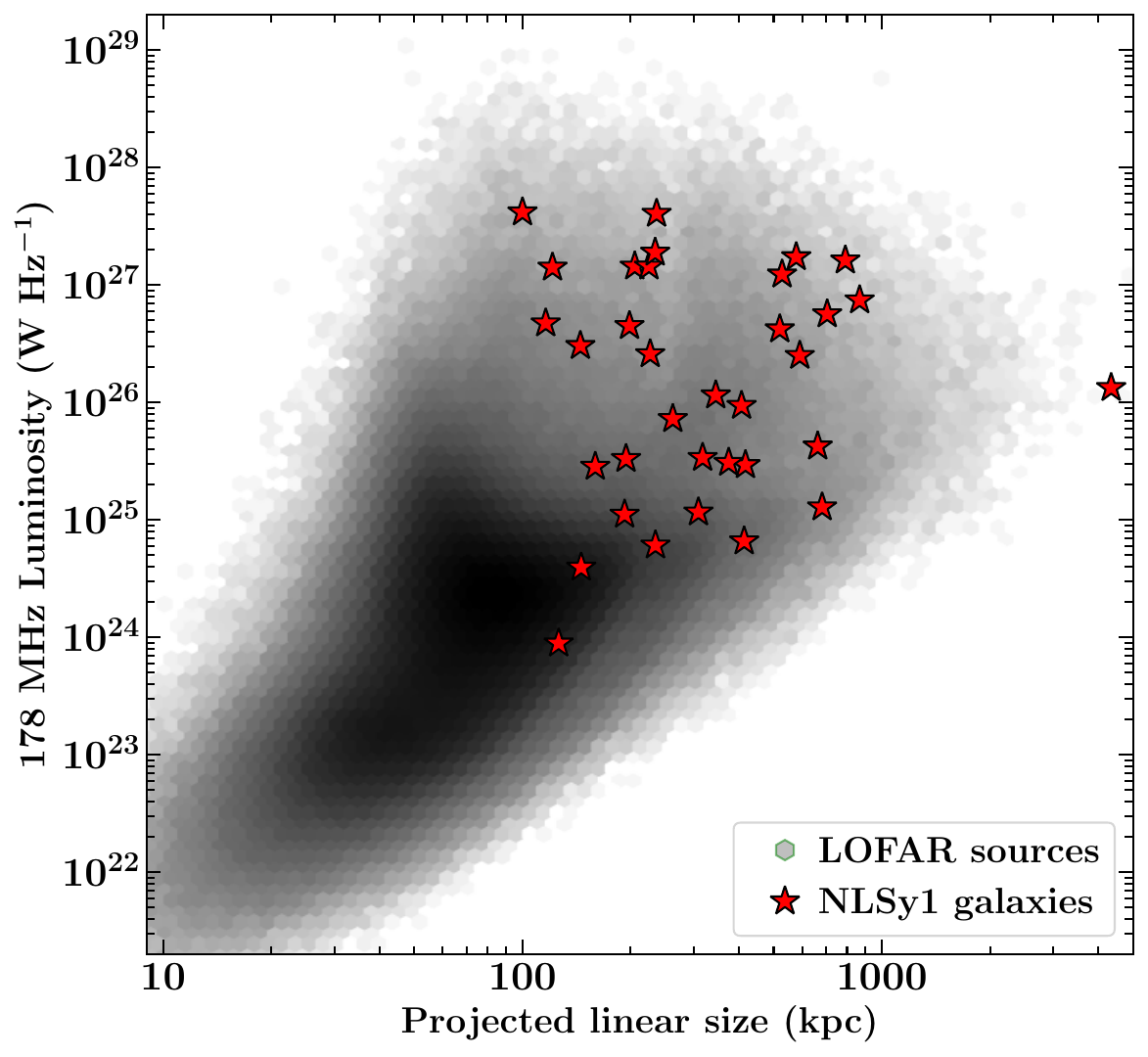}
  \caption{Power vs projected size (P--D diagram) plot of all the double-lobed NLSy1 sources $>$ 100 kpc (listed in Table~\ref{table-2}) overlaid onto LOFAR$-$detected sources published in \citep{Hardcastle2023}.}
  \label{PDdiagram}
\end{figure}

\subsection{Radio luminosity-Projected size diagram}
The radio luminosity-projected size ($P$–$D$) diagram has been traditionally used to study the evolution of radio galaxies of different jet powers \citep{Baldwin1982,Kaiser1997}. In Fig.~\ref{PDdiagram} we show our sample of $>$100 kpc NLSy1s along with LOFAR-detected objects from \citet{Hardcastle2023}. For this analysis, the extended flux densities were calculated as $\mathrm{F_{ext} = F_{total} - F_{core}}$ and extrapolated to the rest frame at 178~MHz, assuming a spectral index of $\mathrm{\alpha_{ext} = -0.8}$.  Almost all these sources have a luminosity $> 10^{25}~W/Hz$, and span the entire range of sizes from 100 kpc to about a Mpc, with J1318+2626 being an outlier with a projected size of about 4.3 Mpc. They lie in the upper region of the $P$–$D$ diagram of the LOFAR sources, suggesting that these follow a similar evolutionary path as other radio galaxies of similar jet power. 

\section{Conclusions} \label{sec:conclusions}
In this paper we report the discovery of 33 NLSy1 galaxies, with a projected linear size of at least 100 kpc. These large-sized radio-emitting NLSy1 galaxies were found by cross-correlating the largest sample of 22,656 NLSy1 galaxies with the FIRST survey and then by examining their VLASS, LOFAR and RACS cutout images. In addition, J1030+5516 which is also in our sample was reported earlier \citep{Rakshit2018}, giving us a sample of 34 sources. The NLSy1 galaxy J0354$-$1340 with a size $>$100 kpc \citep{Vietri2022} is outside the footprint of the SDSS survey, and hence did not appear in our sample. We briefly summarize and comment on the main results here.

    \begin{enumerate}[noitemsep, topsep=0pt]
        \item This work has increased the number of known galaxies with a projected linear size of at least 100 kpc. This has been possible due to the use of a number of sensitive radio surveys, especially LOFAR and RACS.
        
        \item Earlier studies of NLSy1 galaxies with extended structure suggested that they may occur more frequently in FR~II sources. However, we find that the majority of our sources belong to the FR~I category. This is also due to the use of the LOFAR survey data which is more sensitive to diffuse extended emission.

        \item The radio loudness parameter of the $>$100 kpc NLSy1s is $>$100 for $\sim$80 percent of the sources, suggesting that selecting very radio-loud objects could help identify more NLSy1s hosting large-scale jets. 

        \item Four of the NLSy1s are GRGs, with J1318+2626 being the largest with $\rm D_{proj}$=4345 kpc.
        
        \item The radio core dominance parameter of the samples ranged from $-$1.6 to 1.2 with a median at $-$0.1, and $\rm C_D$ is observed to be relatively higher in FR~I objects, consistent with earlier studies. 

        \item A small number of NLSy1s were also classified as blazars, and some of them have also been detected in the $\gamma-$ray band. We find that these sources tend to have more dominant cores and flat radio spectral indices, as expected due to their smaller angles of inclination. However these do not tend to have smaller black hole masses. On the other hand, the GRGs which are expected to be inclined at large angles of inclination, have smaller mass black holes. This suggests that orientation is not playing a dominant role in the estimation of black hole mass, contrary to an earlier suggestion.

        \item The $>$100 kpc NLSy1s lie on a similar region as other luminous radio galaxies in the luminosity-size ($P-D$) diagram, suggesting that these may also
        have a similar evolutionary history.        
    \end{enumerate}

    Upcoming sensitive surveys, especially at low radio frequencies are likely to further increase significantly the number of large $>$100 kpc NLSy1s. This would be useful for detailed multi-wavelength studies of their properties and for understanding the nature, formation, and evolution of this interesting class of AGN.
    
\section{Acknowledgements}
We thank an anonymous reviewer for a careful reading of the manuscript and several valuable comments which have helped improve the paper.

The National Radio Astronomy Observatory is a facility of the National Science Foundation operated under cooperative agreement by Associated Universities, Inc. CIRADA is funded by a grant from the Canada Foundation for Innovation 2017 Innovation Fund (Project 35999), as well as by the Provinces of Ontario, British Columbia, Alberta, Manitoba and Quebec.

LOFAR is the Low Frequency Array designed and constructed by ASTRON. It has observing, data processing, and data storage facilities in several countries, which are owned by various parties (each with their own funding sources), and which are collectively operated by the ILT foundation under a joint scientific policy. The ILT resources have benefited from the following recent major funding sources: CNRS-INSU, Observatoire de Paris and Université d'Orléans, France; BMBF, MIWF-NRW, MPG, Germany; Science Foundation Ireland (SFI), Department of Business, Enterprise and Innovation (DBEI), Ireland; NWO, The Netherlands; The Science and Technology Facilities Council, UK; Ministry of Science and Higher Education, Poland; The Istituto Nazionale di Astrofisica (INAF), Italy.

 This paper includes archived data obtained through the CSIRO ASKAP Science Data Archive, CASDA (https://data.csiro.au).

 The Pan-STARRS1 Surveys (PS1) and the PS1 public science archive have been made possible through contributions by the Institute for Astronomy, the University of Hawaii, the Pan-STARRS Project Office, the Max-Planck Society and its participating institutes, the Max Planck Institute for Astronomy, Heidelberg and the Max Planck Institute for Extraterrestrial Physics, Garching, The Johns Hopkins University, Durham University, the University of Edinburgh, the Queen's University Belfast, the Harvard-Smithsonian Center for Astrophysics, the Las Cumbres Observatory Global Telescope Network Incorporated, the National Central University of Taiwan, the Space Telescope Science Institute, the National Aeronautics and Space Administration under Grant No. NNX08AR22G issued through the Planetary Science Division of the NASA Science Mission Directorate, the National Science Foundation Grant No. AST-1238877, the University of Maryland, Eotvos Lorand University (ELTE), the Los Alamos National Laboratory, and the Gordon and Betty Moore Foundation.

\bibliographystyle{aasjournal}  
\bibliography{ref}  

\end{document}